\documentclass[aps,prd,,showpacs,twocolumn,nofootinbib]{revtex4-1}

\usepackage{amsmath}
\usepackage{amssymb}
\usepackage{graphicx}

\begin{document}


\title{Dynamics of Brans-Dicke cosmology with varying mass fermions}


\author{Dao-Jun Liu}
\email[]{djliu@shnu.edu.cn}
\affiliation{Center for Astrophysics, Shanghai Normal University, 100 Guilin Road, Shanghai 200234,China}


\date{\today}

\begin{abstract}
In this paper,  the cosmological  dynamics of Brans-Dicke theory in which there are fermions with a coupling to BD scalar field as well as a self-interaction potential is investigated. The conditions that there exists a solution which is stable and represents a late-time accelerated expansion of the universe are found. The variable mass of fermions can not vanish exactly during the evolution of the universe once it exists initially.  It is shown that the late-time acceleration depends completely on the self-interaction of the fermion field if our investigation is restricted to the theory with positive BD parameter $\omega$. Provided a negative $\omega$ is allowed, there will be another two class of stable solutions describing late-time accelerated expansion of the universe.
\end{abstract}


\maketitle


\section{Introduction}

Strong evidences from the current cosmological observations such as supernovae type Ia(SNeIa)\cite{A.G. Riess}, cosmic microwave background (CMB)\cite{D. N. Spergel et al.} and large scale structure(LSS)\cite{M. Tegmark et al.} converge upon the fact that the universe is spatially flat and there exists exotic component, dubbed dark energy, which drives the speed-up expansion of the universe. Many scenarios have been proposed to explain the acceleration in the framework of general relativity (GR). The preferred and simplest candidate for dark energy is the Einstein's cosmological constant which can fit the observations well. However, it suffers from the so-called fine-tuning problem  and coincidence problem. On the other hand, the observations are not yet able to confirm that dark energy is indeed a constant. Actually,  many dynamical models of dark energy have been studied extensively during the past over ten years, such as quintessence \cite{Quintessence}, phantom \cite{Phantom}, quintom \cite{Quintom} (see also Ref.\cite{cai} for a detailed review),  tachyon \cite{tacyon}, Chaplygin gas \cite{cg}\footnote{It has been shown that only the generalized version of the Chaplygin gas model is compatible with the observations (see e.g. Ref. \cite{gcg}).}, etc (see Ref.\cite{copeland:2006ej} for a comprehensive review on dark energy models).

GR is a well tested theory in solar system, but it is poorly tested on cosmic scales. Therefore, it is interesting to ask whether the cosmic acceleration is caused by a modified theory of gravity. As is pointed out, the dark energy problem may be essentially an issue of quantum gravity \cite{Witten:2002}. Although a complete  theory of quantum gravity has not been established, some valuable ideas are thought to be the features of the theory of quantum gravity, for example, extra dimensions. Many fundamental theories would induce scalar fields in the usual 4-dimensional space-time, and these scalar fields are in general non-minimal coupled. It is truly remarkable to find that a candidate scalar field of the desired nature is provided by some fundamental theory\footnote{Note that string theory is a quite special type of scalar-tensor theory of gravity, and unless one introduces non-perturbative effects, the theory has a  runaway problem. }. On the other hand, in recent years, the interest in scalar-tensor theories of gravity as viable alternatives to GR is renewed. In particular, some authors have resorted to scalar-tensor theory in order to explain the present accelerating expansion of the universe \cite{Bertolami:2000,Brans-Dicke:DE}.  As the simplest example and prototype of scalar-tensor theory of gravity, Brans-Dicke theory (BD) formulates the gravitational phenomena through the interplay between the metric tensor and a scalar field $\phi$ that controls the intensity of the gravitational constant $G$, while the coupling between $\phi$ and matter is absent. Although there is no direct support from fundamental theories, it is  hard to deny that it appears to provide a small window through which one can look into phenomenological aspects of fundamental theories.

Recently, the possibility that fermion fields as gravitational sources could be responsible for accelerated periods during the expansion of the universe is considered in the framework of GR\cite{Ribas:2005vr} as well as in BD theory of gravity\cite{Samojeden:2010rs}. In the context of cosmology, due to the isotropy and homogeneity of the geometry of the universe, the fermion field configurations should be constant in space. It is shown that the fermion field could behave as dark energy for late-time universe. In this work, we would like to investigate the  dynamics of BD cosmology in which there exists fermions with coupling to BD scalar field as well as a self-interaction potential, and the main focus is looking for the conditions under which there is a solution that is stable and represents a late-time accelerated expansion of the universe.

 This paper is organized
as follows: In section \ref{sec2}, we briefly introduce the tetrad formalism of fermions with variable mass in Brans-Dicke theory of gravity. Then, the cosmological dynamics of the model is investigated in detail in section \ref{sec3}. Finally, in section \ref{sec:conclusion}  we summarize our results and give some discussions.

\section{Fermions with variable mass in Brans-Dicke theory}
\label{sec2}
We start from the Dirac Lagrangian density in Minkowski space-time
\begin{equation}
L_D=\frac{i}{2}\left[\bar{\psi}\gamma^a\partial_a\psi-(\partial_a\bar{\psi})\gamma^{a}\psi \right]-m\bar{\psi}\psi-V,
\end{equation}
where the index $a=0,1,2,3$, $\gamma^{a}$ is the Dirac-Pauli matrices, $m$ is the fermion mass, $\bar{\psi}=\psi^{\dag}\gamma^0$ denotes the adjoint spinor field and $V$, an exclusive function of $\psi$ and $\bar{\psi}$, is the potential representing a fermion self-interaction.

When gravity is taken into account, the generally covariant Dirac Lagrangian becomes
\begin{equation}
L_D=\frac{i}{2}\left[\bar{\psi}\Gamma^{\mu}D_{\mu}\psi-(D_{\mu}\bar{\psi})\Gamma^{\mu}\psi \right]-m\bar{\psi}\psi-V,
\end{equation}
where $\Gamma^{\mu}=e^{\mu}_{a}\gamma^a$ is generalized Dirac-Pauli matrices, $D_{\mu}$ denotes covariant derivatives, which is defined by
\begin{equation}
D_{\mu}=\partial_{\mu}+\frac{g_{\mu\nu}}{4}\left[\Gamma^{\nu}_{\sigma\lambda}-e^{\nu}_{b}(\partial_{\sigma}e^{b}_{\lambda})\right]
 \gamma^{\sigma}\gamma^{\lambda},
\end{equation}
where $\Gamma^{\nu}_{\sigma\lambda}$ and $e^{\nu}_{b}$ denote the Christoffel symbol and the tetrad, respectively. The metric tensor $g_{\mu\nu}$ satisfies the relation
\begin{equation}
g_{\mu\nu}=e^a_{\mu}e^b_{\nu}\eta_{ab},
\end{equation}
where $\eta_{ab}$ is the Minkowski metric tensor.

In the canonical frame, the Lagrangian for the Brans-Dicke theory can be written as
\begin{equation}
L_{G}=-\frac{\phi^2}{8\omega}R +\frac{1}{2}g^{\mu\nu}\nabla _{\mu}\phi\nabla _{\nu}\phi,
\end{equation}
where $R$ is the scalar curvature and $\phi$ is the Brans-Dicke scalar field. $\phi^2$ in the non-minimal coupling term  acts as an effective gravitational constant $G_{\mathrm{eff}}$ in such a way that $G_{\mathrm{eff}}^{-1}=2\pi \phi^2/\omega$. We assume that, just as the Brans-Dicke scalar field $\phi$ adjusts the strength of gravity $G$, the fermion mass $m$ is replaced by $\beta \phi$, with $\beta$ being a dimensionless coupling constant.
Then the Lagrangian for the spinor field with variable mass can be written as
\begin{equation}
L_D=\frac{i}{2}\left[\bar{\psi}\Gamma^{\mu}D_{\mu}\psi-(D_{\mu}\bar{\psi})\Gamma^{\mu}\psi \right]-\beta\phi\bar{\psi}\psi-V,
\end{equation}
So we will take the following total action
\begin{equation}\label{action}
    S=\int d^4x\sqrt{-g}(L_D+L_G+L_M),
\end{equation}
where $L_M$ denotes the Lagrangian of matter fields.

Varying action (\ref{action}) with respect to the spinor field, we obtain
the Dirac equations for the spinor field and its adjoint field
\begin{equation}\label{Dirac1}
    i\Gamma^{\mu}D_{\mu}\psi-\beta \phi\psi-\frac{\partial V}{\partial \bar{\psi}}=0,
\end{equation}
\begin{equation}\label{Dirac2}
    iD_{\mu}\bar{\psi}\Gamma^{\mu}+\beta \phi\bar{\psi}+\frac{\partial V}{\partial {\psi}}=0.
\end{equation}
Similarly, the Einstein equations for Brans-Dicke gravitational dynamics can also be derived from action (\ref{action}),
\begin{widetext}
\begin{equation}\label{EinteinEQ}
\frac{\phi^2}{4\omega}\left(R_{\mu\nu}-\frac{1}{2}g_{\mu\nu}R\right)
+\frac{1}{2\omega}\left(g_{\mu\nu}\phi\Box\phi+g_{\mu\nu}(\nabla\phi)^2-\phi\nabla _{\mu}\nabla _{\nu}\phi-\nabla _{\mu}\phi\nabla _{\nu}\phi\right)
=T^{\phi}_{\mu\nu}+T^D_{\mu\nu}+T^m_{\mu\nu}
\end{equation}
\end{widetext}
where $\Box\equiv g^{\mu\nu}\nabla_\mu\nabla_\nu$ is a covariant D'Lambertian for a scalar field and
\begin{equation}\label{T:phi}
    T^{\phi}_{\mu\nu}=\nabla _{\mu}\phi\nabla _{\nu}\phi-\frac{1}{2}g_{\mu\nu}(\nabla\phi)^2,
\end{equation}
\begin{eqnarray}\label{T:D}
    T^D_{\mu\nu}&=&\frac{i}{4}\left[\bar{\psi}\Gamma_{\mu}D_{\nu}\psi+\bar{\psi}\Gamma_{\nu}D_{\mu}\psi-(D_{\mu}\bar{\psi})\Gamma_{\nu}\psi -(D_{\nu}\bar{\psi})\Gamma_{\mu}\psi \right]\nonumber\\
    &-&g_{\mu\nu}L_D.
\end{eqnarray}
The energy momentum tensor for the matter fields is defined as usual, $T^m_{\mu\nu}=\frac{2}{\sqrt{-g}}\frac{\delta(\sqrt{-g}L_m)}{\delta g^{\mu\nu}}$. In cosmology, it is often be expressed as the form of perfect fluid
\begin{equation}\label{T:Matter}
    T^m_{\mu\nu}=(\rho_m+p_m)u_{\mu}u_{\nu}-p_m g_{\mu\nu},
\end{equation}
where $\rho_m$ and $p_m$ are energy density and pressure of the matter, respectively. The four velocity vector $u_\mu$ normalized as $u_{\mu} u^{\mu}=1$.
Finally, we vary action (\ref{action}) with respect to $\phi$, obtaining the equation of motion for the Brans-Dicke scalar field
\begin{equation}\label{EOM:BD}
    \Box\phi+\frac{\phi R}{4\omega}+\beta \bar{\psi}\psi=0.
\end{equation}

\section{Dynamics of fermion  field in Brans-Dicke cosmology}\label{sec3}
For a spatially flat homogeneous and isotropic universe, the space-time interval is written as usual as
\begin{equation}\label{FRWmetric}
    ds^2=dt^2-a^2(t)(dx^2+dy^2+dz^2),
\end{equation}
where $a(t)$ is the cosmic scale factor. Therefore, the components of tetrad becomes
\begin{equation}\label{tetradComponents}
    e^{\mu}_0=\delta^{\mu}_0, \;\;\;\; e^{\mu}_i=\frac{1}{a(t)}\delta^{\mu}_i
\end{equation}
and the Dirac matrices turn out to be
\begin{equation}
    \Gamma^0=\gamma^0, \;\;\;\; \Gamma^i=\frac{1}{a(t)}\gamma^i,
\end{equation}
from which the covariant derivatives is obtained
\begin{equation}
    D_0=\partial_0,\;\;\;\; D_i=\partial_i-\frac{1}{2}\dot{a}(t)\gamma^i\gamma^0,
\end{equation}
where and thereafter overdot denotes derivative with respect to cosmic time $t$.

Then, the Dirac equations (\ref{Dirac1}) and (\ref{Dirac2}) read\footnote{Just as mentioned in the introduction, for the isotropic and homogeneous universe, the fermion field is an exclusive function of time.}
\begin{equation}\label{Dirac1a}
    \dot{\psi}+\frac{3}{2}H\psi+i\beta\gamma^0 \phi\psi+i\gamma^0\frac{\partial V}{\partial \bar{\psi}}=0,
\end{equation}
\begin{equation}\label{Dirac2a}
    \dot{\bar{\psi}}+\frac{3}{2}H\bar{\psi}-i\beta\gamma^0 \phi\bar{\psi}-i\gamma^0\frac{\partial V}{\partial {\psi}}=0,
\end{equation}
and the  equations of motion for the metric and the Brans-Dicke scalar field become
\begin{equation}\label{feqn1}
\frac{3}{4\omega}\phi^2H^2+\frac{3}{2\omega}H\dot{\phi}\phi=\frac{1}{2}\dot{\phi}^2+\rho_D
+\rho_m
\end{equation}
\begin{equation}\label{feqn2}
\frac{\phi^2}{4\omega}\left(2\frac{\ddot{a}}{a}+H^2\right)
+\frac{1}{\omega}H\dot{\phi}\phi+\frac{1}{2\omega}\ddot{\phi}\phi
+\frac{1}{2\omega}\dot{\phi}^2
=-p_D-\frac{1}{2}\dot{\phi}^2-p_{m}
\end{equation}
\begin{equation}\label{feqn3}
\ddot{\phi}+3H\dot{\phi}+\beta\bar{\psi}\psi-\frac{3}{2\omega}
\left(\frac{\ddot{a}}{a}+H^2\right)\phi=0
\end{equation}
where  $H\equiv \dot{a}/a$ is Hubble parameter and
\begin{equation}\label{rho:D}
    \rho_D\equiv \left(T^D\right)^0_0=\beta\phi\bar{\psi}\psi+V,
\end{equation}
\begin{equation}\label{p:D}
    p_D\equiv -\left(T^D\right)^1_1=\frac{\partial V}{\partial \psi}\frac{\psi}{2}+\frac{\bar{\psi}}{2}\frac{\partial V}{\partial \bar{\psi}} -V.
\end{equation}
Note that, among the equations (\ref{feqn1})-(\ref{feqn3}), only two of them are independent and the rest one can
be derived from Bianchi identity.
From Eqs.(\ref{feqn1})-(\ref{feqn3}), we obtain
\begin{equation}\label{FRW}
    H^2=\frac{2\omega}{3}\frac{\dot{\phi}^2}{\phi^2}
    -2H\frac{\dot{\phi}}{\phi}+\frac{4\omega}{3\phi^2}\left(V+\beta\phi\bar{\psi}\psi+\rho_m\right),
\end{equation}
\begin{equation}\label{dda}
    \frac{\ddot{a}}{a}=\frac{2\omega}{3}\left(\frac{\ddot{\phi}}{\phi}-\frac{\dot{\phi}^2}{\phi^2}\right)
    +2(1+\omega)H\frac{\dot{\phi}}{\phi}-\frac{4\omega}{3\phi^2}(V+\rho_m)-\frac{2\omega\beta\bar{\psi}\psi}{3\phi^2}
\end{equation}

and
\begin{eqnarray}\label{EoM:phi}
    \frac{\ddot{\phi}}{\phi}+3H\frac{\dot{\phi}}{\phi}+\frac{\dot{\phi}^2}{\phi^2}&=&
    \frac{2\omega}{(3+2\omega)\phi^2}\left[\rho_m-3p_m+4V\right.\nonumber\\
    &&-3\left(\frac{\partial V}{\partial \psi}\frac{\psi}{2}+\left.\frac{\bar{\psi}}{2}\frac{\partial V}{\partial \bar{\psi}}\right)\right].
\end{eqnarray}

 We shall consider a pressureless dust matter field, that is, $p_m=0$.
Then, from conservation equation for the energy density of the matter constituent, $\dot{\rho}_m+3H(\rho_m+p_m)=0$, we can obtain that $\rho_m=\rho_{m0}a^{-3}$. In order to analyze the cosmological dynamics, we should select a specific  potential $V$. According to the Pauli-Fierz theorem, $V$ can be an exclusive function of the scalar invariant $(\bar{\psi}\psi)^2$\cite{Ribas:2005vr}. In this work, we choose
\begin{equation}\label{potential}
    V=\lambda(\bar{\psi}\psi)^{2\alpha}
\end{equation}
as in Ref.\cite{Samojeden:2010rs}, where $\lambda$ is a constant and $\alpha$ is a real number.
 The bilinear $\bar{\psi}\psi$ is scalar, hereafter we call it $f$. From Dirac equations (\ref{Dirac1a}) and (\ref{Dirac2a}), it is not difficult to find that $f=f_0a^{-3}$, where $f_0$ is present value of $f$.

Let us define some new dimensionless variables  $y=\frac{\dot{\phi}}{H\phi}$, $\Omega_V=\frac{4\omega V}{3\phi^2H^2}$, $\Omega_{f}=\frac{4\omega \beta f}{3\phi H^2}$ and $\Omega_{m}=\frac{4\omega \rho_{m}}{3\phi^2 H^2}$  as functions of $x=\ln a$.    Therefore,  Friedmann equation (\ref{FRW}) becomes a constraint equation
\begin{equation}\label{constraintEQ}
    1=\frac{2\omega}{3}y^2-2y+\Omega_V+\Omega_f+\Omega_m.
\end{equation}

After a lengthy but straightforward calculations, from Eqs.(\ref{FRW})-(\ref{EoM:phi}), it is shown that the deceleration parameter $q\equiv -\frac{\ddot{a}}{aH^2}$ can be indicated as
\begin{equation}\label{q}
    q=1+\frac{2\omega}{3}y^2-\frac{\Omega_f}{2}
    -\frac{\omega}{3+2\omega}\Omega_m-\frac{\omega(4-6\alpha)}{3+2\omega}\Omega_V,
\end{equation}
where the evolution of $y$ satisfies
\begin{equation}\label{prime:y}
    y'=yq-2y^2-2y+\frac{3}{6+4\omega}\left[\Omega_m+(4-6\alpha)\Omega_V\right],
\end{equation}
where the prime denotes derivative with respect to $x$.

On the other hand, according to the definitions of $\Omega_f$ and $\Omega_m$, as well as the evolutions of $f$ and $\rho_m$, it is easy to find that
\begin{equation}\label{prime:OmegaBeta}
    \Omega_{f}'=-\left(1+y-2q\right)\Omega_{f},
\end{equation}
\begin{equation}\label{prime:OmegaM}
    \Omega_m'=-\left(1+2y-2q\right)\Omega_m.
\end{equation}

The equations (\ref{prime:y})-(\ref{prime:OmegaM}) constitute an autonomous system and we shall use phase space method to investigate the properties of its solutions qualitatively.
 The critical points of the system can be obtained by setting $\Omega_f'=0$, $\Omega_m'=0$ and $y'=0$.
 In Table \ref{table},  all the critical points of the system are listed. The coordinate of the point D is independent of $\omega$ and $\alpha$, while the points B$_\pm$ and C is independent of $\alpha$. As for the coordinates of the other two points A and E, they rely on both of the two parameters $\omega$ and $\alpha$. Noting that merging occurs for the points D and E for $\alpha=\frac{1}{2}$. All the critical points except C are associated with the dust matter vanishing universe. From Eqs.(\ref{prime:OmegaBeta}) and (\ref{prime:OmegaM}), it is obvious that plane  $y-\Omega_m$ ($\Omega_f=0$) and  plane $y-\Omega_f$ ($\Omega_m=0$) are two invariant sub-manifold of the $3$-dimensional phase space. This means that there is no orbit in the phase space for which $\Omega_m$ or $\Omega_f$ can be exactly zero. That is to say, if the initial value of $f=\bar{\psi}\psi$ or coupling constant $\beta$   in the definition of $\Omega_f$ is set to be zero, \textit{i.e.}, the fermion field is initially  massless, it will remain massless. The values of $\Omega_V$ and deceleration parameter at the different  critical points $\Omega_{Vc}$ and $q_c$ are also listed in Table \ref{table}.

What is more interesting is the stable solution which represents that the universe will undergoes a late-time accelerated expansion phase. To investigate the stability of these critical points, we can write the variables near these points $(\Omega_{fc},\Omega_{mc},y_c )$
in the form $y=y_c+\eta_1$, $\Omega_f=\Omega_{fc}+\eta_2$ and $\Omega_{m}=\Omega_{mc}+\eta_3$ with $\eta_i\; (i=1,2,3)$, the perturbations of the variables around the critical points to the first order. This leads to the following equation \begin{equation}\label{Uprime}
    \mathbf{U'}=\mathbf{A}\cdot\mathbf{U},
\end{equation}
where the $3$-column vector $\mathbf{U}=(\eta_i)^{T},\;i=1,2,3$ represents the perturbations of the variables and $\mathbf{A}$ is a constant $3\times 3$ matrix. For stability we require all $3$ eigenvalues of $\mathbf{A}$ to be negative.

\begin{table*}
  \centering
  \begin{tabular}{c|c|c|c}
  \hline
  Points & Coordinate  $(\Omega_{fc},\Omega_{mc},y_c )$& $\Omega_{Vc}$ & $q_c$   \\
   \hline
  A & $\left(0,0,\frac{2-3\alpha}{2\omega(1-\alpha)+1}\right)$  & $\frac{(3+2 \omega ) \left(5+6 \omega +6 \alpha ^2 \omega -6 \alpha  (1+2 \omega )\right)}{3 (1-2 (\alpha-1 ) \omega )^2}$ & $\frac{\left(2-8 \alpha +6 \alpha ^2\right) \omega -1}{2 (\alpha-1 ) \omega -1}$  \\
  \hline
  B$_{\pm}$ & $\left(0,0,\frac{3}{2\omega}\left(1\pm\sqrt{1+\frac{2\omega}{3}}\right)\right)$ &
  $0$& $\frac{3+2 \omega \pm \sqrt{9+6 \omega }}{\omega }$ \\
  \hline
  C & $\left(0, \frac{(2\omega+3)(3\omega+4)}{6(1+\omega)^2}, \frac{1}{2\omega+2} \right)$ & $0$& $\frac{2+\omega }{2+2 \omega }$ \\
  \hline
  D & $(1,0,0)$ & $0 $& $\frac{1}{2}$ \\
  \hline
  E & {\footnotesize $ \left(6 \alpha-2 -12 \omega +36 \alpha  \omega -24 \alpha ^2 \omega,0,3-6\alpha\right)$}  &
  $3 (1-2 \alpha) (3+2 \omega )$& $2-3 \alpha$ \\
  \hline
\end{tabular}
\caption{The critical points of the autonomous system  Eqs.(\ref{prime:y})-(\ref{prime:OmegaM}). The values of $\Omega_V$ and deceleration parameter at the different  critical points $\Omega_{Vc}$ and $q_c$ are also listed.  }\label{table}
\end{table*}

\begin{table*}
  \centering
  \begin{tabular}{c|c|c}
  \hline
Points& Eigenvalues & Stability  \\
   \hline
  A     & {\scriptsize \begin{minipage}{1.5in}\begin{equation*}\begin{aligned}&6 \alpha-3 ,\\
  &\frac{5+6 \omega +6 \alpha ^2 \omega -6 \alpha  (1+2 \omega )}{-1+2 (-1+\alpha ) \omega },\\
  &\frac{1+6 \omega +12 \alpha ^2 \omega -3 \alpha  (1+6 \omega )}{-1+2 (-1+\alpha ) \omega}\end{aligned}\end{equation*}\end{minipage}}
  &{\tiny
  \begin{minipage}{3.5in}\begin{equation*}\begin{aligned}
   & \mathrm{attractor\;\;for\;\;}\alpha <\frac{1}{2}\;\&\left(\omega <\frac{-5+6 \alpha }{6-12 \alpha +6 \alpha ^2}\mathrm{\;or\;}\omega >\frac{-1+3 \alpha }{6-18 \alpha +12 \alpha ^2}\right);\\
   &\mathrm{repeller\;\;for\;\;}\left(\frac{2}{3}<\alpha <1\;\&\;\frac{-1+3 \alpha }{6-18 \alpha +12 \alpha ^2}<\omega <\frac{-5+6 \alpha }{6-12 \alpha +6 \alpha ^2}\right)\\
   &\mathrm{\;\;\;\;\;\;\;\;\;\;\;\;\;\;\;\;\;\;\;\;or\;\;}\alpha \geq 1\;\&\;\left(\omega <\frac{-1+3 \alpha }{6-18 \alpha +12 \alpha ^2}\mathrm{\;\;or\;\;}\omega >\frac{-5+6 \alpha }{6-12 \alpha +6 \alpha ^2}\right);\\
   &\mathrm{saddle\;\;\;\; for\;\; other \;\;cases.}
   \end{aligned}\end{equation*}\end{minipage} } \\
  \hline
  B$_+$ &\raisebox{-1.6ex}[0pt][12pt]{ {\footnotesize \begin{minipage}{1.5in}{\begin{equation*}\begin{aligned}
  &\frac{3+3 \omega \pm \sqrt{9+6 \omega }}{\omega },\\
  &\frac{3 \left(3+2 \omega \pm \sqrt{9+6 \omega }\right)}{2 \omega },\\
  &\frac{3-6 (\alpha-1 ) \omega \pm \sqrt{9+6 \omega }}{\omega }\end{aligned}\end{equation*}}\end{minipage}} }
    &{\tiny
  \begin{minipage}{3.5in}\begin{equation*}\begin{aligned}
   & \mathrm{attractor\;\;for\;\;}\left(\alpha \leq \frac{1}{2}\;\&\frac{-5+6 \alpha }{6-12 \alpha +6 \alpha ^2}<\omega <0\right)\mathrm{or}\left(\alpha >\frac{1}{2}\;\&-\frac{4}{3}<\omega <0\right);\\
   &\mathrm{repeller\;\;for\;\;}(\alpha \leq 1\;\&\;\omega >0)\mathrm{\;\;or\;\;}\left(\alpha >1\;\&\;0<\omega <\frac{-5+6 \alpha }{6-12 \alpha +6 \alpha ^2}\right);\\
   &\mathrm{saddle\;\;\;\; for\;\; other \;\;cases.}
   \end{aligned}\end{equation*}\end{minipage} }
    \\
     \cline{1-1} \cline{3-3}
  B$_-$ &
  &{\tiny  \begin{minipage}{3.5in}\begin{equation*}\begin{aligned}
      &\mathrm{repeller\;\;for\;\;}\omega>-\frac{3}{2}\;\&\;\alpha<\frac{3+6 \omega -\sqrt{9+6 \omega }}{6 \omega };\\
   &\mathrm{saddle\;\;\;\; for\;\; other \;\;cases.}
   \end{aligned}\end{equation*}\end{minipage} }\\
    \hline
  C &$3-6 \alpha ,-\frac{4+3 \omega }{2+2 \omega },\frac{1}{2+2 \omega }$
  &{\tiny  \begin{minipage}{3.5in}\begin{equation*}\begin{aligned}
      & \mathrm{attractor\;\;for\;\;}\alpha >\frac{1}{2}\;\&\;\omega <-\frac{4}{3};\\
   &\mathrm{saddle\;\;\;\; for\;\; other \;\;cases.}
   \end{aligned}\end{equation*}\end{minipage} }\\
  \hline
  D &  $-\frac{3}{2},0,3-6 \alpha $ & \scriptsize{ attractor for $\alpha \geq \frac{1}{2}$; saddle for $\alpha < \frac{1}{2}$.}  \\
  \hline
  E &
  \begin{minipage}{1.5in}\begin{equation*}\begin{aligned}
  & {\footnotesize 6 \alpha-3 ,3 \alpha-2 \pm\sqrt{ A}}
 \end{aligned}\end{equation*}{\tiny  \begin{equation*}\begin{aligned}\textrm{where}\;\; A&=8+24 \omega-8 \alpha  (4+15 \omega )\\ &  -96 \alpha ^3 \omega +3 \alpha ^2 (11+64 \omega )\end{aligned}\end{equation*}}
   \end{minipage}
  &{\tiny
  \begin{minipage}{3.5in}\begin{equation*}\begin{aligned}
   & \mathrm{attractor\;\;for\;\;}\alpha \leq \frac{1}{2}\;\&\frac{8-32 \alpha +33 \alpha ^2}{96 \alpha ^3-192 \alpha ^2+120 \alpha-24 }\leq \omega \leq \frac{3 \alpha-1 }{6-18 \alpha +12 \alpha ^2};\\
   &\mathrm{repeller\;\;for\;\;}\frac{2}{3}\leq\alpha <1\;\&\;\frac{8-32 \alpha +33 \alpha ^2}{96 \alpha ^3-192 \alpha ^2+120 \alpha-24 }\leq \omega \leq \frac{3 \alpha-1 }{6-18 \alpha +12 \alpha ^2}\\
   &\mathrm{\;\;\;\;\;\;\;\;\;\;\;\;\;\;\;\;\;\;\;\;or\;\;}\alpha \geq 1\;\&\;\frac{3 \alpha-1 }{6-18 \alpha +12 \alpha ^2}\leq \omega \leq \frac{8-32 \alpha +33 \alpha ^2}{96 \alpha ^3-192 \alpha ^2+120 \alpha-24 };\\
   &\mathrm{saddle\;\;\;\; for\;\; other \;\;cases.}
   \end{aligned}\end{equation*}\end{minipage} }
\\
  \hline
\end{tabular}
\caption{The eigenvalues  of the linear perturbation matrix and stability for all the critical points.}\label{table2}
\end{table*}

\begin{table*}
  \centering
  \begin{tabular}{c|c}
  \hline
   Points& Conditions for a stable and late-time acceleration solution  \\
   \hline
  A      &  \begin{minipage}{2.8in}{\scriptsize \begin{equation*}\begin{aligned} &\;\;\;\;\;\alpha <\frac{1}{3}\;\&\; (\omega <\frac{-5+6 \alpha }{6-12 \alpha +6 \alpha ^2}\|\omega >\frac{1}{2-8 \alpha +6 \alpha ^2})\\
  &\mathrm{or}\;\;\alpha =\frac{1}{3}\;\&\;\omega <-\frac{9}{8}\\
  &\mathrm{or}\;\;\frac{1}{3}<\alpha <\frac{1}{2}\;\&\frac{1}{2-8 \alpha +6 \alpha ^2}<\omega <\frac{-5+6 \alpha }{6-12 \alpha +6 \alpha ^2} \end{aligned}\end{equation*}}\end{minipage} \\
  \hline
  B$_+$
  &{\footnotesize $-\frac{4}{3}<\omega <0\;\&\;\alpha >\frac{1+2 \omega +\sqrt{1+\frac{2 \omega }{3}}}{2\omega} $}
    \\
 \hline
 C & {\footnotesize$-2<\omega <-\frac{4}{3}\;\&\;\alpha >\frac{1}{2}$} \\
  \hline
  D  &{\footnotesize False}  \\
  \hline
  E  & {\footnotesize False}\\
  \hline
\end{tabular}
\caption{The conditions for the critical points representing  a stable and late-time accelerated solution. }\label{table3}
\end{table*}

%

The eigenvalues of the linear perturbation matrix $\mathbf{A}$ for the critical points and the stability properties are summarized  in table \ref{table2}, while the conditions for a stable and late-time acceleration solution are listed in table \ref{table3}. It is found that all the critical points but B$_-$ can be stable for some values of parameters $\omega$ and $\alpha$. However, the stable solutions presented by D and E can not describe a universe that undergoes a late-time acceleration. Furthermore, it is worth noting that only point A allows BD parameter $\omega>0$ for a stable and late-time accelerated solution.

\begin{figure}
\centering
  \includegraphics[width=8cm]{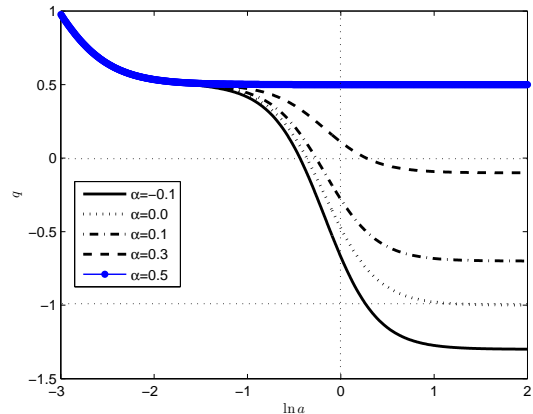}\\
  \caption{The evolution of deceleration parameter $q$ with respect to $\ln a$ for different values of $\alpha$, where  the BD parameter is chosen to be $\omega=5\times 10^4$, and initial values of $y$, $\Omega_f$ and $\Omega_m$ is set at present ($a=1$) to be $-0.00001$, $0.05$ and $0.3$, respectively.}\label{fig1}
\end{figure}

\begin{figure}
\centering
  \includegraphics[width=8cm]{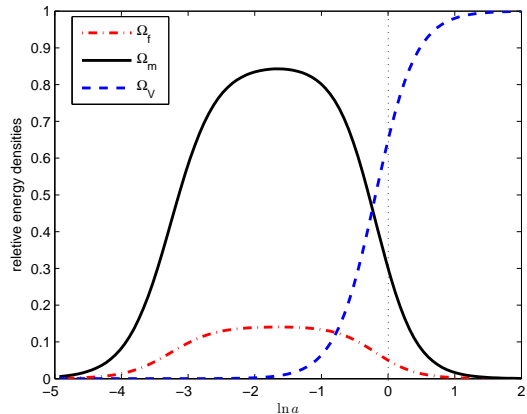}\\
  \caption{The evolution of relative energy densities with respect to $\ln a$, where $\alpha$ and $\omega$ is respectively chosen to be $-0.05$ and $5\times 10^4$ and the initial (present) values of $y$, $\Omega_f$ and $\Omega_m$ is set by $-0.00001$, $0.05$ and $0.3$.}\label{fig2}
\end{figure}

As is well known, in the limit case $\omega\rightarrow \infty$, the standard Einstein's GR is recovered from the BD theory, except for some special cases, for example,  in which the trace of the matter energy-momentum tensor vanishes \cite{Faraoni:1999a}. And the solar-system experiments constraint severely that the value of $\omega$ obeys $\omega >4\times 10^{4}$ \cite{Bertotti:2003b}. With this understanding, the late-time acceleration of expansion of the universe will be entirely dependent upon the self-interaction of the fermion field as is illustrated in figure \ref{fig1}, where the BD parameter is chosen to be $\omega=5\times 10^4$, and initial values of $y$, $\Omega_f$ and $\Omega_m$ is set at present ($a=1$) to be $-0.00001$, $0.05$ and $0.3$, respectively. We recall that in the standard FRW universe containing only  single barotropic fluid in the framework of GR,  the deceleration parameter $q$ and the equation of state of barotropic fluid  $w=p/\rho$ have a relation $2q=1+3w$. Therefore, from figure \ref{fig1}, we can find that the equation of state of effective dark energy is able to cross the phantom divide $w=-1$ for some values of $\alpha$ and the big rip may also occur in the future. In figure \ref{fig2}, we plot the evolution trajectories of relative energy densities $\Omega_m$, $\Omega_f$ and $\Omega_V$. It is not surprising that the energy density of the fermion field will eventually overcome the energy density of the matter field and this occurs roughly at the time the universe going into an accelerated expansion phase, while the energy density contributed by the variable mass of fermion field keeps similar pace with that of the matter field.

\begin{figure}
\centering
  \includegraphics[width=8cm]{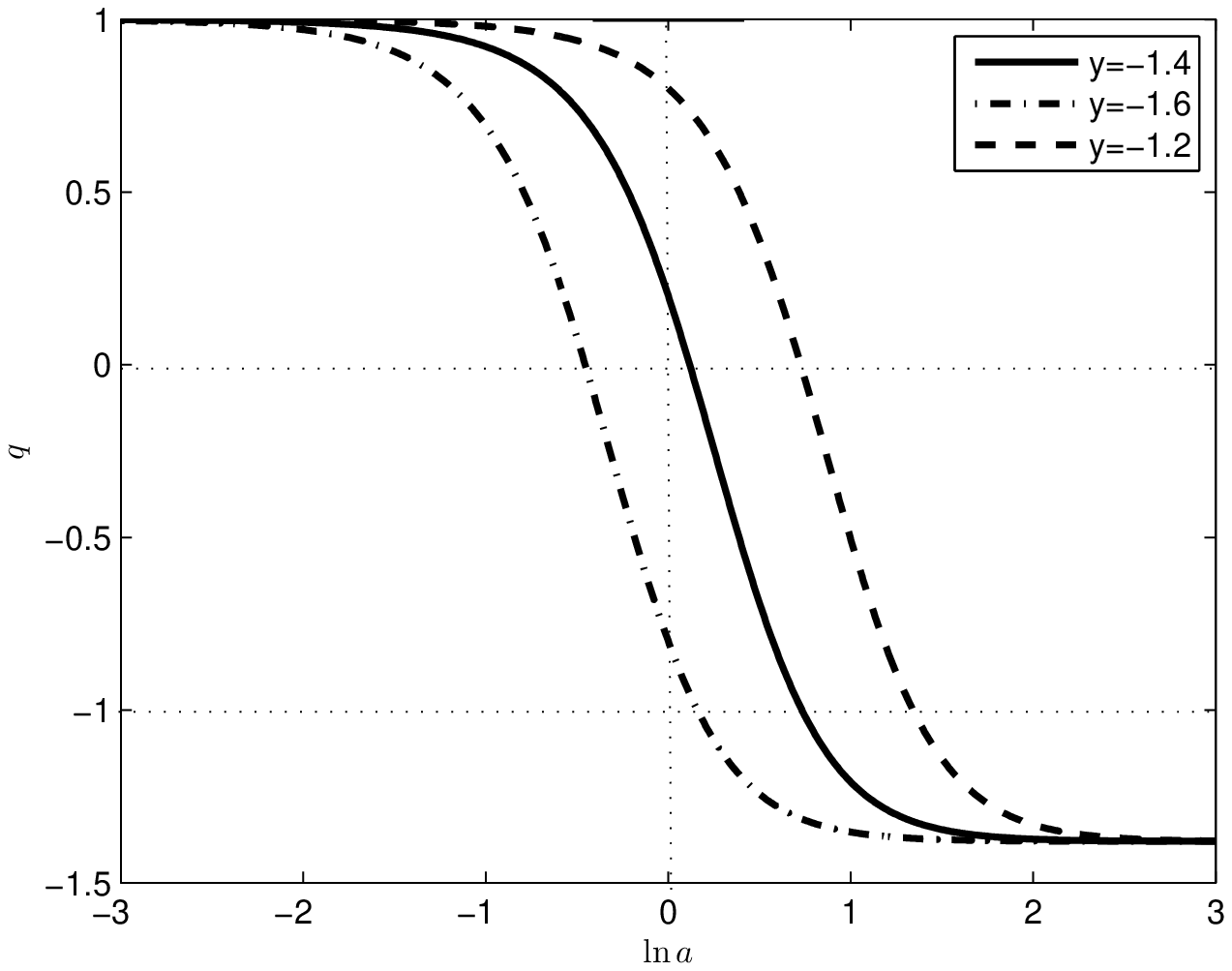}\\
  \caption{The evolution trajectories of deceleration parameter $q$ with respect to $\ln a$ for different initial (present) values of $y$, where  $\alpha$ and $\omega$ is respectively chosen to be $1$ and $-1.25$. }\label{fig:bqx}
\end{figure}

\begin{figure}
\centering
  \includegraphics[width=8cm]{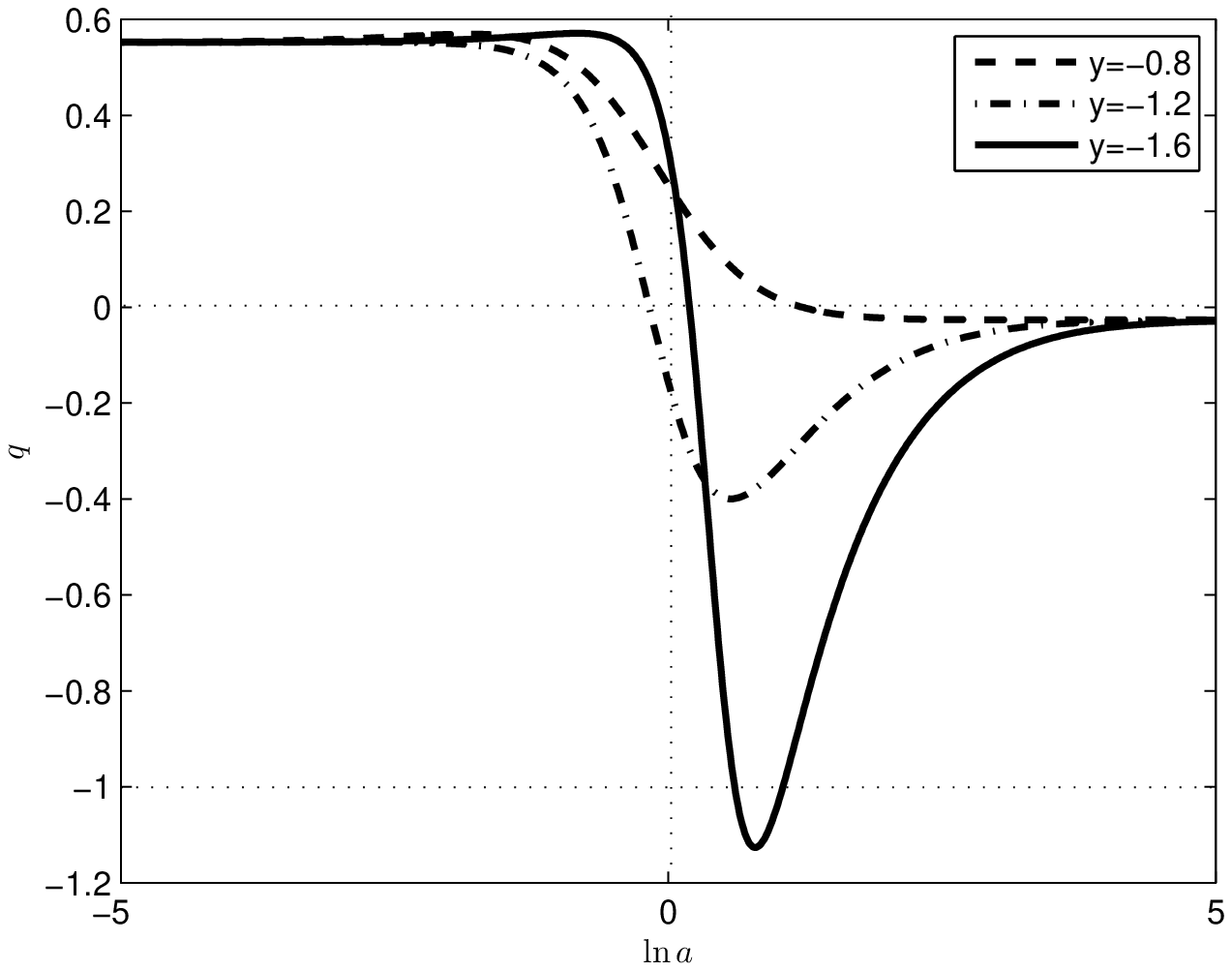}\\
  \caption{The evolution trajectories of deceleration parameter $q$ with respect to $\ln a$ for different initial (present) values of $y$, where  $\alpha$ and $\omega$ is respectively chosen to be $0.9$ and $-1.95$.}\label{fig:cqx}
\end{figure}

\begin{figure}
\centering
  \includegraphics[width=8cm]{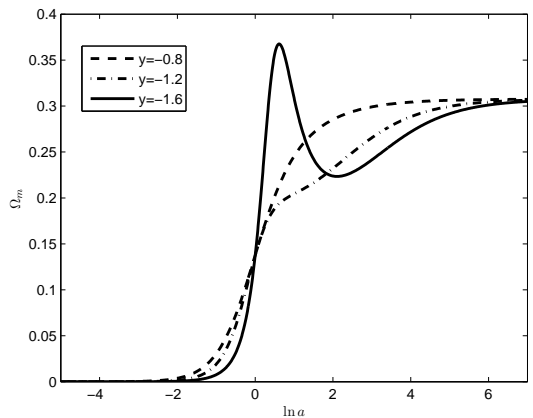}\\
  \caption{The evolution trajectories of relative energy density of dust matter with respect to $\ln a$ for different initial (present) values of $y$, where  $\alpha$ and $\omega$ is respectively chosen to be $0.9$ and $-1.95$. }\label{fig:cOm}
\end{figure}

On the other hand, if we allow the BD parameter $\omega$ to be less than zero on the cosmological scales, there will exist another two class of stable late-time accelerated expansion solutions which are represented by critical points B$_+$ and C and illustrated in figure \ref{fig:bqx} and figure \ref{fig:cqx}, respectively. These solutions do not determined by the self-interaction of the fermion field, because in these two class of solutions $\Omega_V$ converges towards zero and the parameter $\alpha>1/2$. The  solutions illustrated in figure \ref{fig:bqx} denote a phantom field dominated universe which will suffer from the fate of big rip. However, the solutions illustrated in figure \ref{fig:cqx} will avoid this violent doom,   although the equation of state of effective dark energy under certain conditions may become less than $-1$ within some period of time. More interestingly, it is shown in figure \ref{fig:cOm} that, in this class of solutions, the relative energy density of dust matter will not vanish but converge to a reasonable finite value.

\section{Conclusions}\label{sec:conclusion}
Brans-Dicke theory is a natural alternative and a simple extension of general relativity. In this work, we have studied the cosmological dynamics of Brans-Dicke theory in which there are fermions with a coupling to BD scalar field as well as a self-interaction potential. Because of the isotropy and homogeneity of the geometry of the universe, the fermion field is independent of space and becomes an exclusive function of time.  The conditions that there exist solutions which are stable and describe a late-time accelerated expansion of the universe are found. The variable mass of fermions can not vanish exactly during the evolution of the universe once it obtains initially.  It is shown that the late-time acceleration depends completely on the self-interaction of the fermion field if our investigation is restricted to the theory with positive BD parameter $\omega$. Provided a negative $\omega$ is allowed, there will be another two class of stable solutions describing late-time accelerated expansion of the universe which are independent of the self-interaction of fermion field.

The strongest constraint to date on the Brans-Dicke theory of gravity has been put on the solar system scale\cite{Bertotti:2003b}, however, it should be pointed out that such constraint does not necessarily apply on scales much larger than those of the measurements, and on epochs much different from the present. The solar system experiments only probe scales in gravitational equilibrium, where the background expansion of the universe is negligible, therefore, they would not reveal spatial or time variation of the gravitational constant on larger scales. Through a certain  mechanism, the BD parameter $\omega$ becomes variable from small scales to large scales.  It is even not excluded that on the largest scales the gravitational constant becomes effectively negative. Some indications for a negative $\omega$ (or effectively negative gravitational constant) have  been obtained \cite{Bertolami:2000}. Therefore, it is interesting to investigate such a model presented here in the context of a detailed viable gravitational theory with variable effective $\omega$, for example, the abnormally weighting energy theory \cite{AWE}.

\begin{acknowledgments}
This work is supported in part by National Natural Science Foundation of China under Grant  No. 10503002, Shanghai Commission of Science and technology under Grant No. 06QA14039 and Innovation Program of Shanghai Municipal Education Commission under Grant No. 09YZ148.
\end{acknowledgments}


\end{document}